\documentclass[aps,twocolumn,showpacs]{revtex4}
\usepackage{graphicx}
\usepackage{amsfonts}
\usepackage{amssymb}
\usepackage{amsbsy}
\usepackage{amsmath}
\usepackage{mathrsfs}
\usepackage{latexsym}
\usepackage{natbib}
\usepackage{bm}
\usepackage{subfigure}
\usepackage{color}
\usepackage{wasysym}
\usepackage{esvect}
\usepackage{wrapfig}
\usepackage{lipsum}
\usepackage{bbm, dsfont}

\def\varphit{\tilde{\varphi}}
\def\pit{\tilde{\pi}}

\def\Tr{\mathrm{Tr}\hspace{1pt}}

\begin{document}

\title{Perturbative approach to the entanglement entropy and the area law in 
Fock and polymer quantization}

\author{Subhajit Barman}
\email{sb12ip007@iiserkol.ac.in}

\author{Gopal Sardar}
\email{gopal1109@iiserkol.ac.in}

\affiliation{ Department of Physical Sciences, 
Indian Institute of Science Education and Research Kolkata,
Mohanpur - 741 246, WB, India }


\date{\today}

\begin{abstract}

The area dependence of entanglement entropy of a free scalar field is often 
understood in terms of coupled harmonic oscillators. In Schrodinger 
quantization, the Gaussian nature of ground state wave-function for these 
oscillators is sufficient to provide the exact form of the reduced density 
matrix and its eigenvalues, thus giving the entanglement entropy. However, in 
polymer quantization, the ground state is not Gaussian and the formalism which 
can provide the exact analytical form of the reduced density matrix is not yet 
known. In order to address this issue, here we treat the interaction between two 
coupled harmonic oscillators in the perturbative approach and evaluate the 
entanglement entropy in Schrodinger and polymer quantization. In contrary to 
Schrodinger quantization, we show that in high frequency regime the entanglement 
entropy decreases for polymer quantization keeping the ratio of coupling 
strength to the square of individual oscillator frequency fixed. Furthermore, 
for a free scalar field, we validate the area dependence of entanglement entropy 
in Fock quantization and demonstrate that polymer quantization produces a 
similar area law.

\end{abstract}

\maketitle

\section{Introduction}

The fact that one can incorporate thermodynamical attributes to a black hole was 
first introduced in the seminal work of Bekenstein \cite{Bekenstein:1972tm, 
Bekenstein:1973ur}. In these articles and others \cite{Bekenstein:1974ax, 
Kallosh:1992wa, hawking1975, Bekenstein:1975tw, Hawking:1976de, Gibbons:1976pt, 
Davies:1978mf, Israel:2003kv, Ashtekar:1997yu, Mukohyama:1997aq, Bardeen:1973gs} 
the authors demonstrated that intrinsic entropy $\mathcal{S}_{BH}$ of a black 
hole should be proportional to the area $\mathcal{A}_{h}$ of its event horizon  
$\mathcal{S}_{BH} = \frac{1}{4}M_{Pl}^2 \mathcal{A}_{h}$, where $M_{Pl}$ is the 
Planck mass. Then the natural question appeared is how to connect the concept of 
quantum states to this entropy of event horizon \cite{Frolov:1993ym, 
Barvinsky:1994jca, Kiefer:1998mi, Strominger:1996sh} as horizon is not different 
than any other classical surface with no special local dynamics. To answer this 
question and to provide a more general realization of the entropy associated to 
a black hole the authors in \cite{Bombelli:1986rw, Srednicki:1993im} presented 
the idea in terms of entanglement entropy. Here it is shown that entanglement 
entropy of a free scalar field in a certain spatial region is proportional to 
its area. In these articles the reduced density matrix, essential for estimating 
the entanglement entropy, is obtained by tracing over the spatial degrees of 
freedom of the ground state density matrix residing inside the considered 
region.

In the regular formulation of entanglement entropy estimation 
\cite{Srednicki:1993im} firstly the scalar field is partially Fourier 
transformed with respect to the angular coordinates. The resulting Fourier field 
Hamiltonian is still dependent on the radial coordinate and it is discretized by 
assuming a lattice of finite size and inter-atomic spacing. This discretization 
transforms the Fourier Hamiltonian to be a collection of coupled harmonic 
oscillators. The ground state wave-function for these coupled harmonic 
oscillators then provide the corresponding ground state density matrix for the 
field. Subsequently using the Gaussian nature of this ground state wave-function 
the reduced density matrix and its eigenvalues are obtained which would produce 
the entanglement entropy. However this Gaussian nature is a feature specific to 
the Schrodinger quantization. In polymer quantization \cite{Ashtekar:2002sn, 
Halvorson-2004-35, Hossain:2010eb}, the quantization method used in loop quantum 
gravity \cite{Ashtekar:2004eh, Rovelli2004quantum, Thiemann2007modern}, the 
ground state wave-functions are expressed in terms of Mathieu functions. Using 
these polymer wave-functions it is still unknown how to obtain the analytic form 
of reduced density matrix.

In this article we consider a perturbative approach to circumvent these 
difficulties and obtain the entanglement entropy for free scalar field using 
Fock and polymer quantization. We treat the interaction between coupled harmonic 
oscillators in perturbative manner to get the related ground state and 
eigen-values of the reduced density matrix. Firstly we use this procedure to 
evaluate the entanglement entropy for two coupled harmonic oscillators in 
Schrodinger and polymer quantization. Then by considering free scalar field we 
obtain the area law in Fock quantization. Furthermore we apply polymer 
quantization in this formulation and verify that the field theoretic 
entanglement entropy obeys a similar area law.

In section \ref{EE_usual} we briefly review the procedures to derive the 
entanglement entropy in usual formulation. In this section the detailed 
description of the considered system is given. Following in section 
\ref{EE_perturbed} we recall the perturbative formulation and construct the 
framework to estimate the entanglement entropy utilizing this technique. In the 
subsequent sections we use this formulation to obtain entanglement entropy for 
two coupled harmonic oscillators in Schrodinger and polymer quantization. 
Following parts include the realization of the area law of entanglement entropy 
in Fock and polymer quantization utilizing perturbative formulation. We argue on 
the implications of the obtained results and conclude with the discussion.

\section{Entanglement entropy and the area law}\label{EE_usual}

In the standard derivations of entanglement entropy \cite{Bombelli:1986rw, 
Srednicki:1993im, Das:2008sy, Das:2005ah, Ahmadi:2005mh, Das:2007mj, Das:2007sj, 
Muller:1995mz, Das:2006vx, Jonker2016EntanglementEO} one considers a system of 
coupled harmonic oscillators as a basis. In particular the eigen-values of the 
reduced density matrix for two coupled oscillators give the entanglement entropy 
corresponding to a single oscillator. These eigenvalues are used for a set of 
coupled harmonic oscillators, which are obtained from the discretized 
Hamiltonian of a free scalar field, to get the area law of entanglement entropy. 
In this section we briefly review the key aspects of these procedures and the 
considered systems, which will also be useful to construct the perturbative 
formulation.

\subsection{Entanglement entropy for two coupled harmonic oscillators}

In order to understand entropy from entanglement at first a system of two 
coupled harmonic oscillators \cite{Srednicki:1993im, Das:2008sy} is considered. 
The two unit mass oscillators are denoted by their position and momentum 
$(x_1,p_1)$ and $(x_2,p_2)$. The total system can be described by the 
Hamiltonian
\begin{eqnarray}
  H &=& \frac{1}{2}\left[p_1^2 + p_2^2+\omega_0^2 (x_1^2+ 
x_2^2)+k_1^2(x_1-x_2)^2 \right] \label{eq:hamiltonian-full1} \nonumber \\
&=& \frac{1}{2}\left[p_+^2 + \omega_+^2 x_+^2\right] + \frac{1}{2}\left[p_-^2 + 
\omega_-^2 x_-^2\right] ~,
\end{eqnarray}
where the normal coordinates $x_\pm=(x_1\pm x_2)/\sqrt{2}$, $p_\pm=(p_1\pm 
p_2)/\sqrt{2}$ and normal frequencies $\omega_+=\omega_0$, 
$\omega_-=(\omega_0^2+2k_1^2)^{1/2}$ are defined to make the Hamiltonian 
decoupled. In decoupled form the ground state wave-function becomes simplified 
and can be expressed in terms of the normal coordinates as
\begin{equation}
 \psi_0(x_1,x_2) = \left( \frac{\omega_+ \omega_-}{\pi^2} 
\right)^{\frac{1}{4}} 
\exp\left[ -\frac{\omega_+ x_+^2 + \omega_- x_-^2}{2}
\right]~.\label{eq:GS-wavefn-normal-freq}
\end{equation}
From the expression (\ref{eq:GS-wavefn-normal-freq}) one can find out the ground 
state density matrix to be $\rho(x_1,x_2;x_1',x_2') = 
\psi_0(x_1,x_2)~\psi_0^{*}(x_1',x_2')$. To discuss about the entanglement 
entropy corresponding to a single oscillator  one needs to find its associated 
reduced density matrix. The reduced density matrix is obtained by tracing out 
the density matrix with respect to the position degree of freedom of a single 
oscillator, expressed as
\begin{equation}
\rho^{r}(x_2,x_2') = \int_{-\infty}^{\infty}dx_1 ~ 
\psi_0(x_1,x_2)~\psi_0^{*}(x_1,x_2')~.\label{eq:reduced-general}
\end{equation}
The reduced density matrix describes whether the system is in mixed or pure 
state and the corresponding entanglement entropy is defined as $S_{E} = 
-\Tr[\rho^{r} \ln\rho^{r}]$. In a suitable basis one can evaluate the 
entanglement entropy by obtaining the eigenvalues of the reduced density matrix. 
In particular for two coupled harmonic oscillators the resulting reduced density 
matrix from equation (\ref{eq:reduced-general}) has eigenvalues
\begin{equation}
\lambda_n = (1-\xi^2)\;\xi^{2n}~~\textup{where}~~\xi=\frac{\sqrt{\omega_-} 
-\sqrt{\omega_+} } { \sqrt{\omega_-} 
+\sqrt{\omega_+}}~.\label{eq:eigenvalue-gen}
\end{equation}
Then the corresponding entanglement entropy \cite{Srednicki:1993im, 
Das:2008sy, Chandran:2018wwc} becomes
\begin{eqnarray}
 S_{E}(\xi) = -\sum_n\lambda_n\ln{\lambda_n}= 
-\ln{(1-\xi^2)}-\frac{\xi^2}{1-\xi^2}\ln{\xi^2}.\label{eq:ent-entropy-2HO-gen}
\end{eqnarray}

\subsection{Entanglement entropy for N-coupled harmonic oscillators}

Now it is important to understand the entanglement entropy corresponding to 
$N-$coupled harmonic oscillators to get the area law of entanglement entropy for 
free scalar field. The general Hamiltonian for $N$ coupled harmonic oscillator 
is
\begin{equation}
 H = \frac{1}{2} \sum_{j=1}^{N} p_{j}^2 + \frac{1}{2} \sum_{j,k=1}^{N} 
x_{j}K_{jk}x_{k}~,\label{eq:general-Ncoupled-HO-hamiltonian}
\end{equation}
where the matrix $K$ describes the potential and interaction. The diagonal 
elements of $K$ give the frequency square of individual oscillator and symmetric 
off diagonal elements provide the interaction between two adjacent oscillators. 
With the help of a suitably chosen orthogonal matrix $U$ this interaction matrix 
is diagonalized to $K_D$ as $K=U^T K_D U$. The ground state wave-function of 
this $N-$coupled harmonic oscillator (\ref{eq:general-Ncoupled-HO-hamiltonian}) 
can be expressed as
\begin{equation}
 \psi_{0}(x_1,...,x_N) = \left(\frac{Det.\Omega}{\pi^N}\right)^{\frac{1}{4}} 
\exp{\left[ -\frac{x.\Omega.x}{2} \right]}~,\label{eq:general-Ncoupled-HO-GS}
\end{equation}
where $\Omega=U^T K^{1/2}_D U$. From this wave-function one can obtain the 
reduced density matrix when first $\mathbbm{n}$ of the total $N$ oscillators are 
traced out \cite{Srednicki:1993im}. The reduced density matrix is further 
evaluated using a general form of the matrix $\Omega$,
\begin{equation}
 \Omega = \left( {\begin{array}{ccccc}
   A & B\\
   B^T & C\\
  \end{array} } \right)~,
\end{equation}
where $A$ is a $\mathbbm{n}\times \mathbbm{n}$ matrix corresponding to the first 
$\mathbbm{n}$ oscillators, $C$ is a $(N-\mathbbm{n})\times(N-\mathbbm{n})$ 
matrix and $B$ is a $\mathbbm{n}\times(N-\mathbbm{n})$ matrix. In terms of few 
newly defined quantities $\beta=(1/2)B^T A^{-1}B$ and $\gamma= C-\beta$ the 
reduced density matrix becomes
\begin{equation}
 \rho_{out}(x,x')\sim \exp{\left[ -(x.\gamma.x+x'.\gamma.x')/2+x.\beta.x' 
\right]}~,\label{eq:Rrho_Nc}
\end{equation}
where $x$ and $x'$ consist of the $(N-\mathbbm{n})$ oscillators after the 
integration over the first $\mathbbm{n}$ degrees of freedom. 
$x=V^T\gamma^{-1/2}_{D}y$ is defined, where $\gamma=V^T\gamma_{D}V$ such that 
$\gamma_{D}$ is diagonal and $V$ is orthogonal. Then one shall get 
$\rho_{out}(x,x')\sim \exp{\left[ -(y.y+y'.y')/2+y.\beta'.y' \right]}$, where 
$\beta'=\gamma^{-1/2}_{D} V\beta V^T \gamma^{-1/2}_{D}$. Now moving to the basis 
$z=W^{T}y$, such that $\beta'$ is diagonalized as  $\beta'_{D}=W^T\beta'W$, one 
gets
\begin{equation}\label{eq:rho_nreduced}
 \rho_{out}(z,z')\sim \prod_{j=\mathbbm{n}+1}^{N}\exp{\left[ 
-(z_{j}^2+{z'}_{j}^{2})/2+\beta'_{j} z_{j}z'_{j} \right]}~,
\end{equation}
where $\beta'_{j}$ are the eigenvalues of $\beta'$. Then the entanglement 
entropy \cite{Srednicki:1993im} corresponding to $(N-\mathbbm{n})$ oscillators 
turns out to be $S=\sum_jS(\xi_{j})$, with $S(\xi)$ given by equation 
(\ref{eq:ent-entropy-2HO-gen}) and $\xi_{j}^{2} = 
\beta'_{j}/[1+(1-{\beta'}_{j}^2)^{1/2}]$.

\subsection{Entanglement entropy for free scalar field and area law}

In order to discuss about the area law for entanglement entropy, a free massive 
scalar field $\Phi(\vv{x})$ is considered with mass $\mu$ and conjugate momenta 
$\Pi(\vv{x})$. In Minkowski spacetime the Hamiltonian \cite{book:Peskin, 
book:ADas, padmanabhan2016quantum} corresponding to the scalar field is
\begin{equation}\label{eq:Field-hamiltonian-massive}
 H = \frac{1}{2}\int d^3x~\left[ 
\Pi^2(\vv{x})+|\nabla\Phi(\vv{x})|^2 + \mu^2 \Phi^2(\vv{x}) \right]~.
\end{equation}
In terms of partial Fourier decomposition the field $\Phi(\vv{x})$ and the 
conjugate momentum $\Pi(\vv{x})$ are transformed with respect to the angular 
coordinates as
\begin{eqnarray}\label{eq:Field-fourier-transform}
 \Phi(\vv{x}) &=& \sum_{l,m} 
Z_{lm}(\theta,\phi)~\frac{\varphi_{lm}(r)}{r}~,\nonumber\\
\Pi(\vv{x}) &=& \sum_{l,m} 
Z_{lm}(\theta,\phi)~\frac{\pi_{lm}(r)}{r}~,
\end{eqnarray}
where $Z_{lm}(\theta,\phi)$ denotes real spherical harmonics and the Fourier 
field modes satisfy a commutation relation $[\varphi_{lm}(r),\pi_{l'm'}(r')] = 
i\delta_{ll'}\delta_{mm'}\delta(r-r')$ among themselves. With this definition of 
field decomposition from Eqn. (\ref{eq:Field-fourier-transform}) the 
Hamiltonian now becomes $H=\sum_{lm}H_{lm}$, where
\begin{eqnarray}\label{eq:Hamiltonian-massive-FT}
H_{lm} &=&  \frac{1}{2}\int_{0}^{\infty} dr \left\{ 
\pi^2_{lm}(r)+r^2\left[\frac{\partial}{\partial 
r}\left(\frac{\varphi_{lm}(r)}{r}\right)\right]^2 
\right.\nonumber\\
~&& ~~~~~~~~~~~~~\left. +\left(\frac{l(l+1)}{r^2}+\mu^2\right) \varphi^2_ {
lm}(r) \right\}~.
\end{eqnarray}
Next the radial coordinate $r$ is discretized forming a lattice with 
inter-atomic spacing $a$ and size $L=(N+1)a$. The inverse of the spacing 
$a^{-1}$ signifies the ultraviolet cutoff while the inverse system size $L^{-1}$ 
denotes infrared cutoff. This discretization makes the Hamiltonian look like a 
set of coupled harmonic oscillators
 \begin{eqnarray}
H_{lm} &=&  \frac{1}{2a} \sum_{j=1}^{N} \left[ 
\pi^2_{lm,j}+\left(j+\frac{1}{2}\right)^2\left(\frac{\varphi_{lm,j}}{j}
-\frac{\varphi_ {lm,j+1}}{j+1} 
\right)^2 \right.\nonumber\\
~&& ~~~~~~~~~~~~~~~~\left. 
+\left\{\frac{l(l+1)}{j^2}+\mu^2a^2\right\}\varphi^2_{
lm,j} \right]~,\label{eq:Hamiltonian-massive-FT-discrete1}
\end{eqnarray}
such that $\varphi_{lm,N+1}=0$ and $[\varphi_{lm,j},\pi_{l'm',j'}] = 
i\delta_{ll'}\delta_{mm'}\delta_{jj'}$. Comparison of this Hamiltonian with the 
Hamiltonian for $N-$coupled harmonic oscillators from equation 
(\ref{eq:general-Ncoupled-HO-hamiltonian}) gives \cite{Das:2008sy}
\begin{eqnarray}
 K_{jk} &=& \frac{1}{j^2} \left[ 
l(l+1)\delta_{jk}+\frac{9}{4}\delta_{j1}\delta_{k1}+\left(N-\frac{1}{2}
\right)^2\delta_{jN}\delta_{kN} \right.\nonumber\\
&+& \left. (\mu a)^2j^2\delta_{jk} +
\left\{\left(j+\frac{1}{2}\right)^2+\left(j-\frac{1}{2}\right)^2\right\}\delta_{
jk(j\ne 1, N)} \right]\nonumber\\
&-&\left[\frac{\left(k+\frac{1}{2}\right)^2}{k\left(k+1\right)}\right]\delta_{j,
 k+ 1} 
-\left[\frac{\left(j+\frac{1}{2}\right)^2}{j\left(j+1\right)}\right]\delta_{j, 
k- 1}~.\label{eq:Interaction_matrix_discrete_hamiltonian}
\end{eqnarray}
The discretization of the radial coordinate enables one to get a finite 
expression of matrix $K$ denoting the potential energy and interaction. This in 
turn would enable one to obtain the entanglement entropy when a finite number 
$\mathbbm{n}$ of spatial points are traced out in total $N+1$ points. Then as 
one plots the entanglement entropy with respect to $(\mathbbm{n}+1/2)^2$, one 
gets a straight line which represents the celebrated area law for entanglement 
entropy \cite{Srednicki:1993im}. We shall present the area curve of entanglement 
entropy coming from perturbative formulation along with the curve obtained from 
this usual formulation together in the next section.

\section{Entanglement entropy in perturbative approach}\label{EE_perturbed}

As discussed in previous section using Gaussian ground state-wave function of 
coupled harmonic oscillators from Schrodinger quantization Eqn. 
(\ref{eq:GS-wavefn-normal-freq}), one can easily evaluate the exact form of 
reduced density matrix in Eqn. (\ref{eq:reduced-general}). However, in polymer 
quantization, a quantization method used in \emph{loop quantum gravity}, the 
ground state wave function is obtained in terms of Mathieu functions. To the 
best of our knowledge evaluation of exact analytical form of reduced density 
matrix is not possible even for two coupled oscillators using these polymer 
wave-functions. This constraint further debars one to obtain the eigen-values 
for $N-$coupled oscillators in polymer quantization and motivates us to take 
help of the perturbation technique.

In this section we are going to apply perturbation to describe entanglement 
entropy of coupled harmonic oscillators. We express the Hamiltonian 
corresponding to the coupled oscillators in terms of a non interacting free 
Hamiltonian $H_{0}$ and a net interaction term $\lambda H_{int}$ as
\begin{equation}\label{eq:Hamiltonian_interaction}
 H = H_0 +\lambda H_{int}~,
\end{equation}
When interaction strength is smaller than the strength of free Hamiltonian, 
which is obtained for small $\lambda$, one can express the ground state 
$|\Omega\rangle$ corresponding to the whole system in a perturbative manner as
\begin{equation}
 |\Omega\rangle=|0\rangle + \lambda |0^{1}\rangle + \lambda ^2 |0^{2}\rangle + 
\dots~,\label{eq:GS-general-perturbation}
\end{equation}
where $|0\rangle$ denotes the ground state corresponding to the non-interacting 
Hamiltonian $H_0$. On the other hand $|0^{1}\rangle$ and $|0^{2}\rangle$ denotes 
the first and second order perturbative corrections to the non-interacting ground 
state. From the time independent perturbation theory \cite{sakurai1985modern, 
griffiths2018introduction, cohen1991quantum} one obtains the first order 
correction to the ground state as
\begin{equation}
 |0^{1}\rangle = \sum_{n\neq 0}  \frac{\langle 
n|\hat{H}_{int}|0\rangle}{E_0-E_n} |n\rangle ~,
\label{eq:perturbative-gen-wavefn1}
\end{equation}
where $E_n$ denotes the energy of the $n^{th}$ excited state corresponding to 
the non interacting Hamiltonian. The second order correction to the ground state 
is expressed as
\begin{eqnarray}
|0^{2}\rangle &=&- \sum_{n\neq 0}  \frac{\langle 
0|\hat{H}_{int}|0\rangle\langle 
n|\hat{H}_{int}|0\rangle}{(E_0-E_n)^2} |n\rangle \nonumber\\
&& +~\sum_{m\neq 0} \sum_{n\neq 0}  \frac{\langle 
m|\hat{H}_{int}|n\rangle\langle 
n|\hat{H}_{int}|0\rangle}{(E_0-E_m)(E_0-E_n)} |m\rangle ~.
\label{eq:perturbative-gen-wavefn2}
\end{eqnarray}
We shall use these perturbative corrections to obtain the actual ground state 
upto certain perturbative order in the system of coupled harmonic oscillators. 
We mention that while discussing polymer quantization \cite{Hossain:2010eb} we 
shall consider only the $\pi-$periodic sector for our calculations. In 
$\pi-$periodic sector, except for the ground state, the even and odd energies 
become degenerate in high energy regimes. Now as we are interested in the ground 
state density matrix it is convenient for us to consider the non-degenerate 
perturbation theory.

\subsection{Entanglement entropy for two coupled harmonic oscillators}

We begin with a system of two coupled harmonic oscillators. We recall the 
Hamiltonian from equation (\ref{eq:hamiltonian-full1}) and observe that it can 
be expressed in form of Eqn. (\ref{eq:Hamiltonian_interaction}) with $H_0= H_1 + 
H_2$ and $\lambda H_{int} = -k_1^2 x_1x_2$, where $H_j = [p_j^2+\omega^2 
x_j^2]/2$  and $\omega=(\omega_{0}^2+k_1^2)^{1/2}$. Perturbative methods can be 
applied when $k_1^2$ is smaller than $\omega^2$, which is always true for 
nonzero $\omega_{0}$. Then in this system of two coupled oscillators the 
correction to the ground state wave-function due to  first order perturbation 
would be
 \begin{eqnarray}
\lambda |0^{1}\rangle = A_{nn}~ |n\rangle_{1} \otimes |n\rangle_{2} = A_{nn}~ 
|n,n\rangle ~,\label{eq:perturbative-wavefn1}
 \end{eqnarray}
where in the second compact notation of the wave-function the first index 
corresponds to first oscillator and the second one corresponds to second 
oscillator. Here the operation of $\hat{x}_j$ on the corresponding ground state 
is given by 
\begin{equation}
 \hat{x}_j |0\rangle_{j} = \sum_nC^j_{0n}|n\rangle_{j}~,\label{eq:operator_x}
\end{equation}
where in general the most dominating term comes from a single excitation 
$|n\rangle_{j}$. Then we get for two coupled oscillators
\begin{equation}\label{eq:perturbative_coefficient}
 A_{nn} = \frac{k_1^2}{E_{n,n}-E_{0,0}}C^1_{0n}C^2_{0n}~.
\end{equation}
Considering up to the $1^{st}$ order perturbation, the normalized ground 
state will be $ |\Omega\rangle=N_{nn}^{1}\left[|0,0\rangle + A_{nn}~ 
|n,n\rangle\right]$, where the normalization factor 
$N_{nn}^{1}=(1+A_{nn}^2)^{-1/2}$. The corresponding reduced density matrix for 
the first oscillator would be
\begin{equation}
 \hat{\rho}_{1} = \Tr_{2}(|\Omega\rangle \langle\Omega|) = 
(N_{nn}^{1})^2\left[ 
|0\rangle \langle 0| + A_{nn}^2 |n\rangle \langle n| \right]~,
\end{equation}
where the states now correspond to the first oscillator. This reduced density 
matrix has eigen-values $(N_{nn}^{1})^2$ and $(N_{nn}^{1}A_{nn})^2$, and it 
would give the entanglement entropy
\begin{equation}
 S_{E}^1 = -(N_{nn}^{1})^2\left[ \ln{\left(N_{nn}^{1}\right)^2} + A_{nn}^2 
\ln{\left(A_{nn}N_{nn}^{1}\right)^2}\right]~.\label{eq:ent-entropy-1st-pert1}
\end{equation}
Now we consider $2^{nd}$ order perturbation and from Eqn. 
(\ref{eq:perturbative-gen-wavefn2}) we observe that the first quantity would 
vanish as $\langle 0|\hat{x}_1\hat{x}_{2}|0\rangle=0$, when discussing two 
coupled oscillators. Then second order correction to ground state can be
expressed as
 \begin{eqnarray}
\lambda^2|0^{2}\rangle = A_{nn}k^2_{1}\sum_{m\neq0}
\frac{\langle m|\hat{x}_1\hat{x}_{2}|n,n\rangle}{E_{m}-E_{0,0}}~|m\rangle
~.\label{eq:perturbative-wavefn2}
 \end{eqnarray}
We shall evaluate this quantity explicitly in Schrodinger quantization and 
compare the qualitative difference of resulting entanglement entropy with the 
result obtained from first order perturbation.

\subsubsection{Schrodinger quantization}


\begin{figure}
  \includegraphics[width=0.8\linewidth]{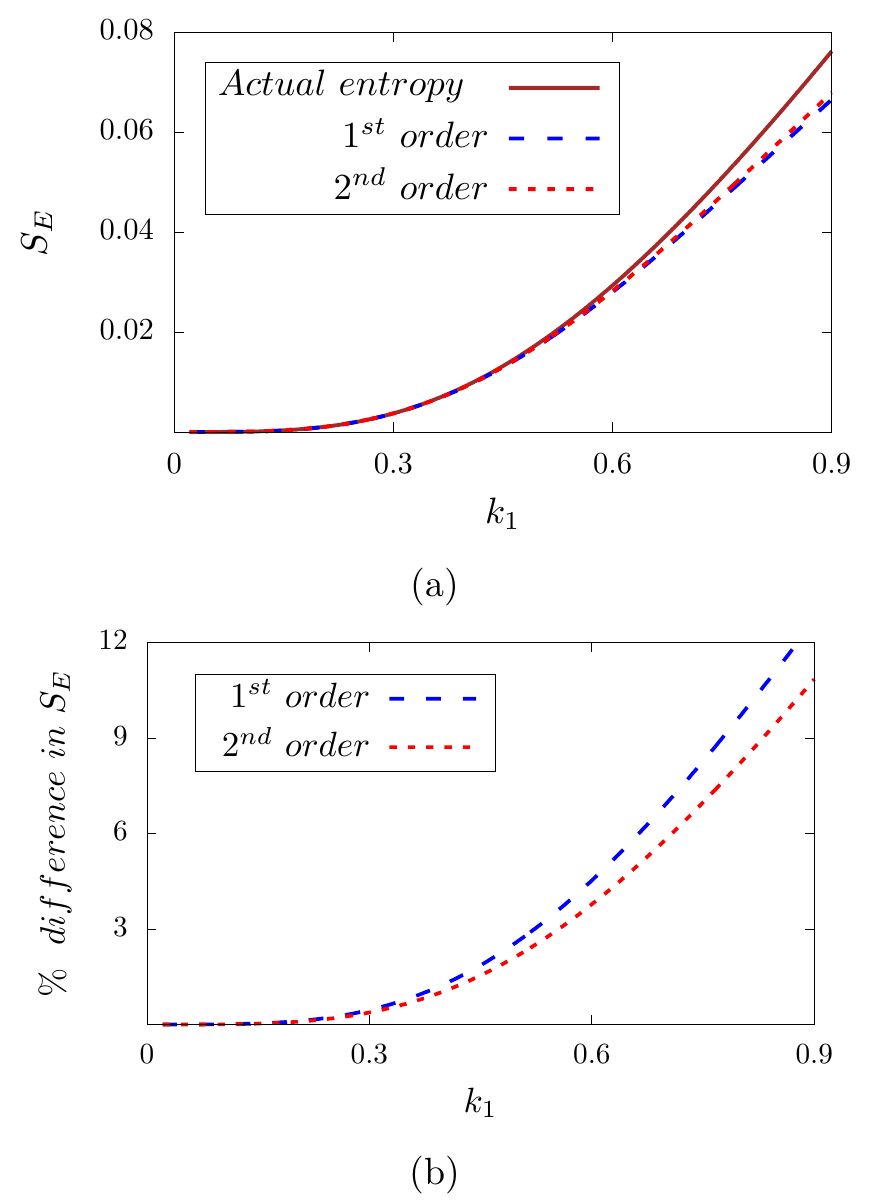}
  \caption{(a) We have plotted the entanglement entropy of two coupled harmonic 
oscillators with respect to varying $k_{1}$ in unit of $\omega_{0}$, from both 
actual and perturbative formulation. (b) The percentage difference of 
perturbative entanglement entropies from actual entropy is plotted, with respect 
to varying $k_{1}$ in unit of $\omega_{0}$.}
  \label{fig:Perturbative-EE-comparison}
\end{figure}

%
In Schrodinger quantization $C^j_{0n}=\delta_{1,n}/\sqrt{2\omega}$ and 
$E_n=(n+1/2)\omega$, then we have $\hat{x}_j|0\rangle_{j} = 
1/\sqrt{2\omega}~|1\rangle_{j}$ and $E_{11}-E_{00}=2\omega$. Then 
$A_{11}=A=k_1^2/4\omega^2$ and $\lambda |0^{1}\rangle = k_1^2/4\omega^2~ 
|1,1\rangle $. The entanglement entropy would be given by Eqn. 
(\ref{eq:ent-entropy-1st-pert1}). When the interaction $k_1^2$ is very small 
compared to the frequency square $\omega^2(\approx\omega_0^2, ~\textup{in~ 
that~ case})$ of the individual oscillator, the expression of entanglement 
entropy from Eqn. (\ref{eq:ent-entropy-1st-pert1}) can be simplified to
\begin{equation}
 S_{E}^1 \approx - \ln{\left[1-\left(\frac{k_1}{2\omega_0}\right)^4\right]} - 
\left(\frac{k_1}{2\omega_0}\right)^4 
\ln{\left[\left(\frac{k_1}{2\omega_0}\right)^4\right]}~.\label{
eq:ent-entropy-1st-pert2 }
\end{equation}
One can observe that in similar conditions an exactly same expression for 
entanglement entropy is obtained from equation (\ref{eq:ent-entropy-2HO-gen}) 
using (\ref{eq:eigenvalue-gen}). Thus at least for two coupled oscillators, when 
the interaction is comparatively much lower than the frequency square, the first 
order perturbation produces reasonable results in accordance with the results 
from actual formulation. This fact can also be verified from FIG. 
\ref{fig:Perturbative-EE-comparison}$:(a)$. Similarly in Schrodinger 
quantization the second order correction to the wave-function from Eqn. 
(\ref{eq:perturbative-wavefn2}) becomes
 \begin{eqnarray}
\lambda^2|0^{2}\rangle = A^2\left[|2,2\rangle + 
\sqrt{2} \left(|0,2\rangle + |2,0\rangle \right)\right] 
~.\label{eq:perturbative-wavefn22}
 \end{eqnarray}
Then the normalized ground state wave-function would be 
\begin{small}
\begin{eqnarray}
 |\Omega\rangle &=& N_{2}\left[|0,0\rangle + A~ |1,1\rangle \right.\nonumber\\
 ~&+& \left. A^2\left\{|2,2\rangle + \sqrt{2} \left(|0,2\rangle + |2,0\rangle 
\right)\right\}\right]~,
\end{eqnarray}
\end{small}
where $N_{2}=(1+A^2+5A^4)^{-1/2}$ is the normalization constant. One obtains 
the reduced density matrix corresponding to the first oscillator as
\begin{eqnarray}
 \hat{\rho}_{1} &=& N_{2}^2\left[ (1+2A^4)|0\rangle \langle 0| + A^2 |1\rangle 
\langle 1| + 3A^4 |2\rangle \langle 2| \right.\nonumber\\
&&~+ ~ \left. \sqrt{2}A^2(1+A^2) \left\{|2\rangle \langle 0| + |0\rangle \langle 
2|\right\} \right]~.\label{eq:reduced-perturb1}
\end{eqnarray}
This reduced density matrix has eigenvalues 
\begin{eqnarray}
\lambda_{1} &=& N_{2}^2A^2~\nonumber\\
\textup{and}\nonumber\\
\lambda_{2,3} &=& \frac{N_{2}^2}{2}\left[ 
1+5A^4\pm\sqrt{(1+A^2)^2(9A^4-2A^2+1)} 
\right]~,\nonumber\\
\end{eqnarray}

which would give the entanglement entropy to be $S_{E} = -\sum_{s=1}^3 
\lambda_s\ln{\lambda_s}$ . This entanglement entropy and the entanglement 
entropy obtained from first order perturbation are plotted with the actual 
entropy in FIG. \ref{fig:Perturbative-EE-comparison}$:(a)$ with respect to a 
varying coupling between the two oscillators. The percentage difference of the 
obtained result using perturbative techniques from the actual entanglement 
entropy is plotted in FIG. \ref{fig:Perturbative-EE-comparison}$:(b)$. From 
these figures we observe that when the coupling is small compared to the 
individual frequency square of the oscillators, perturbation method is quite 
elegant to study entanglement entropy of coupled harmonic oscillators. 
Furthermore from these figures we also observe that the results from second 
order perturbation does not drastically improve compared to first order 
perturbation. On the other hand as our main objective is to understand the 
qualitative nature of entanglement entropy from perturbation, it is expected 
that first order perturbation would be good enough to satisfy our requirement.

\subsubsection{Polymer quantization}

In this part we are going to discuss about entanglement entropy for the system 
of two coupled harmonic oscillators in polymer quantization. Perturbation 
techniques will be used to obtain the entanglement entropy as the wave-functions 
arising from polymer quantization cannot be handled analytically like the 
Gaussian wave-functions. Here we start with a brief overview of the technical 
aspects of polymer quantization.

Polymer quantization \cite{Hossain:2010eb} is a background independent 
quantization procedure arising from loop quantum gravity (LQG). In polymer 
quantization apart form the Planck constant $\hbar$ a new dimension-full 
parameter $\lambda$ is introduced. Here the elementary operators are 
configuration operator $\hat{x}$ and translation operator $\hat{U}_{\lambda} 
\equiv \hat{e^{i\lambda p}}$ and their actions are defined as
\begin{equation}\label{eq:poly_operator}
 \hat{x} e^{i p x_{k}} = i\frac{\partial}{\partial p} e^{i p x_{k}}~,~~~~
 \hat{U}_{\lambda} e^{i p x_{k}} =  e^{i p (x_{k}+\lambda)}~.
\end{equation}
These operators satisfy the basic commutator $[\hat{x}, \hat{U}_{\lambda}] = 
-\lambda \hat{U}_{\lambda}$. Now with the definition of translation operator 
from Eqn. (\ref{eq:poly_operator}) and the inner product
\begin{equation}\label{eq:poly_innerP}
 \langle x|x' \rangle = \lim_{T\to\infty}\frac{1}{2T} \int_{-T}^{T} dp e^{-i p 
x}e^{i p x'} = \delta_{x,x'}~,
\end{equation}
it is observed that a momentum operator cannot be defined as the translation 
operator is not continuous in its parameter. However to describe the kinetic 
energy part of the Hamiltonian one must have a suitable expression of the 
momentum operator. In this case the momentum operator should be $\lambda$ 
dependent and to be given in terms of the translation operator. One simple 
definition of the momentum operator as considered in \cite{Hossain:2010eb} is
\begin{equation}\label{eq:poly_pdef}
 \hat{p}_{\lambda} = 
\frac{1}{2i\lambda}(\hat{U}_{\lambda}-\hat{U}^{\dagger}_{\lambda})~.
\end{equation}
One can then express the eigen-value equation $\hat{H}\psi=E\psi$, where 
$H=p^2/2m+m\omega^2x^2/2$ represents the Hamiltonian corresponding to a simple 
harmonic oscillator with mass $m$, as
\begin{equation}\label{eq:poly_eom}
 \psi''(u) + \left[ \sigma-\frac{1}{2g^2}\cos{(2u)} \right]\psi(u) = 0~,
\end{equation}
which represents a Mathieu equation \cite{Abramowitz1964handbook}. Here 
$\lambda=\lambda_{\star}$, $u=\lambda_{\star}p+\pi/2$, 
$g=m\omega\lambda_{\star}^2$ and $\sigma = 2E/g\omega-1/2g^2$. The above 
differential equation has periodic solutions for $\sigma$ representing the 
Mathieu characteristic value functions
\begin{eqnarray}\label{eq:poly_WF}
 \psi_{2n}(u) &=& \pi^{-1/2} ce_{n}(1/4g^2,u), 
~~~\sigma=\mathcal{A}_{n}(g)\nonumber\\
 \psi_{2n+1}(u) &=& \pi^{-1/2} se_{n+1}(1/4g^2,u), 
~~~\sigma=\mathcal{B}_{n}(g)~.\nonumber\\
\end{eqnarray}
For $(n=0,1,...)$, $ce_{n}$ and $se_{n}$ represent the elliptic cosine and 
sine functions, where for even $n$ they are $\pi-$periodic and for odd $n$ they 
are $\pi-$antiperiodic functions. The corresponding energy eigen values are 
given by 
\begin{eqnarray}\label{eq:poly_energy}
 E_{2n} &=& \frac{\omega}{4g}\left[2g^2\mathcal{A}_{n}(g)+1\right]\nonumber\\
 E_{2n+1} &=& \frac{\omega}{4g}\left[2g^2\mathcal{B}_{n+1}(g)+1\right]~.
\end{eqnarray}
Using the asymptotic expansions of the Mathieu characteristic value functions 
$\mathcal{A}_{n}(g)$ and $\mathcal{B}_{n}(g)$ one can get in small $g$ limit, 
i.e. when $g\ll1$
\begin{eqnarray}\label{eq:poly_energy_gL}
\frac{E_{2n}}{\omega}\approx\frac{E_{2n+1}}{\omega}=\left(n+\frac{1}{2}
\right)-\frac{(2n+1)^2+1}{16}g+\mathcal{O}(g^2)~.\nonumber\\
\end{eqnarray}
On the other hand in high $g$ regimes, i.e. when $g\gg1$, one gets
\begin{eqnarray}\label{eq:poly_energy_gH}
\frac{E_{0}}{\omega}=\frac{1}{4g}+\mathcal{O}\left(\frac{1}{g^3}\right),~\frac{
E_ { 2n-1}}{\omega} \approx\frac{E_{ 2n}}{\omega} = 
\frac{n^2g}{2}+\mathcal{O}\left(\frac{1}{g}\right).\nonumber\\
\end{eqnarray}
From the asymptotic expression (\ref{eq:poly_energy_gL}) we observe that when 
$g\ll1$ the energy levels corresponding to the $\pi-$periodic and 
$\pi-$antiperiodic sectors becomes degenerate among themselves. On the other 
hand from (\ref{eq:poly_energy_gH}) we observe that for $g\gg1$ the energy 
levels within the separate $\pi-$periodic and $\pi-$antiperiodic sectors becomes 
degenerate. We shall consider only the $\pi-$periodic sector of the 
wave-functions, containing the non-degenerate ground state, to discuss about the 
corresponding entanglement entropy. We want to mention that the asymptotic 
expressions of the energy eigen-values from Eqn. (\ref{eq:poly_energy_gL}, 
\ref{eq:poly_energy_gH}) are also utilized in \cite{Hossain:2014fma, 
Hossain:2015xqa, Barman:2017vqx} to observe the Unruh and Hawking effect for 
polymer observer. One can also look into \cite{Kajuri:2015oza, 
Stargen:2017xii, Louko:2017emx, Kajuri:2017zmr} where polymer quantization is 
used in different systems to study particle creation.
\vspace{0.5cm}

\emph{Entanglement entropy in polymer quantization:}
In this part we evaluate the perturbative corrections to the ground state in 
polymer quantization, which basically requires the estimation of $A_{nn}$. The 
operation $\hat{x}_j|0\rangle_j$ is already discussed in Eqn. 
(\ref{eq:operator_x}) and in polymer quantization $C^j_{0n}$ are given by
\begin{eqnarray}
  C^j_{0n} &=& {}_{j}\langle 
n|\hat{x}_{j}|0\rangle_{j}=\lambda_{\star}\int_0^{2\pi} 
\psi^j_n ~\frac{\partial}{\partial u^j}\psi^j_0~ du^j 
~,\label{eq:Cn_general_poly}
\end{eqnarray}
where $u^j=\lambda_{\star}p^j+\pi/2$ and $\lambda_{\star}$ is the polymer length 
scale, see \cite{Hossain:2010eb}. There are infinite number of non-zero 
$C^j_{0n}$ in polymer quantization, where as in Schrodinger quantization there 
is only one $C^j_{0n}=\delta_{1,n}/\sqrt{2\omega}$. In order to compute polymer 
corrections we only consider the first and most dominating non-zero $C^j_{0n}$, 
which is $C^j_{03}$. In small $g=\omega \lambda_{\star}^2$ limit, i.e. when 
$g\ll1$, these coefficients are given by
\begin{equation}
C^j_{03}= C_{03}=\frac{i}{\sqrt{2\omega}}[1-\frac{3}{4} g+\mathcal{O}(g^2)]~,
\end{equation}
and the corresponding energy correction is given by
\begin{equation}
 \Delta 
E_{30}=E_3-E_0=\omega[1-\frac{g}{2}+\mathcal{O}(g^2)]~.
\end{equation}
The expression of $A_{nn}$, now obtained as $A_{33}$, is changed and using Eqn. 
(\ref{eq:perturbative_coefficient}) becomes
\begin{eqnarray}
 A_{33} = A&\approx& -\frac{k_1^2}{4\omega^2(1-g/2)}(1-\frac{3}{2}g)\nonumber\\
 &\approx&-\frac{k_1^2}{4\omega^2}(1-g)~.
\end{eqnarray}
One can observe from this expression of $A$ that as one takes $g\to 0$, one gets 
back the result from Schrodinger quantization. We note that the sign of $A$ do 
not affect the end result as the entanglement entropy is obtained using $A^2$. 
In the ultraviolet limit when $g\gg1$, one has the expressions
\begin{equation}
 C^j_{03}=C_{03} = i\sqrt{\frac{g}{2\omega}}\left[\frac{1}{4g^2} + 
\mathcal{O}\left(\frac{1}{g^6}\right)\right]~,
\end{equation}
and
\begin{equation}
 \Delta 
E_{30}=\omega\left[2g+\mathcal{O}\left(\frac{1}{g^3}\right)\right]~.
\end{equation}
Then the expression of $A_{33}$ is
\begin{equation}
 A_{33}=A\approx -\left(\frac{k_1}{2\omega}\right)^2 \frac{1}{2(2g)^4}~.
\end{equation}
We know that the reduced density matrix in first order perturbation has 
eigenvalues $\lambda_{1}=N_{1}^2=(1+A^2)^{-1}$ and $\lambda_{2} = (N_{1}A)^{2}$. 
Then for fixed $(k_1/\omega)$ as we take $g\to\infty$, we observe that 
$\lambda_{1}\to1$ and $\lambda_{2}\to0$, because in this limit $A\to0$. Then the 
entanglement entropy evaluated from these eigenvalues would vanish, providing a 
very new feature in ultraviolet regime of energy in polymer quantization.


\begin{figure} 
  \includegraphics[width=0.8\linewidth]{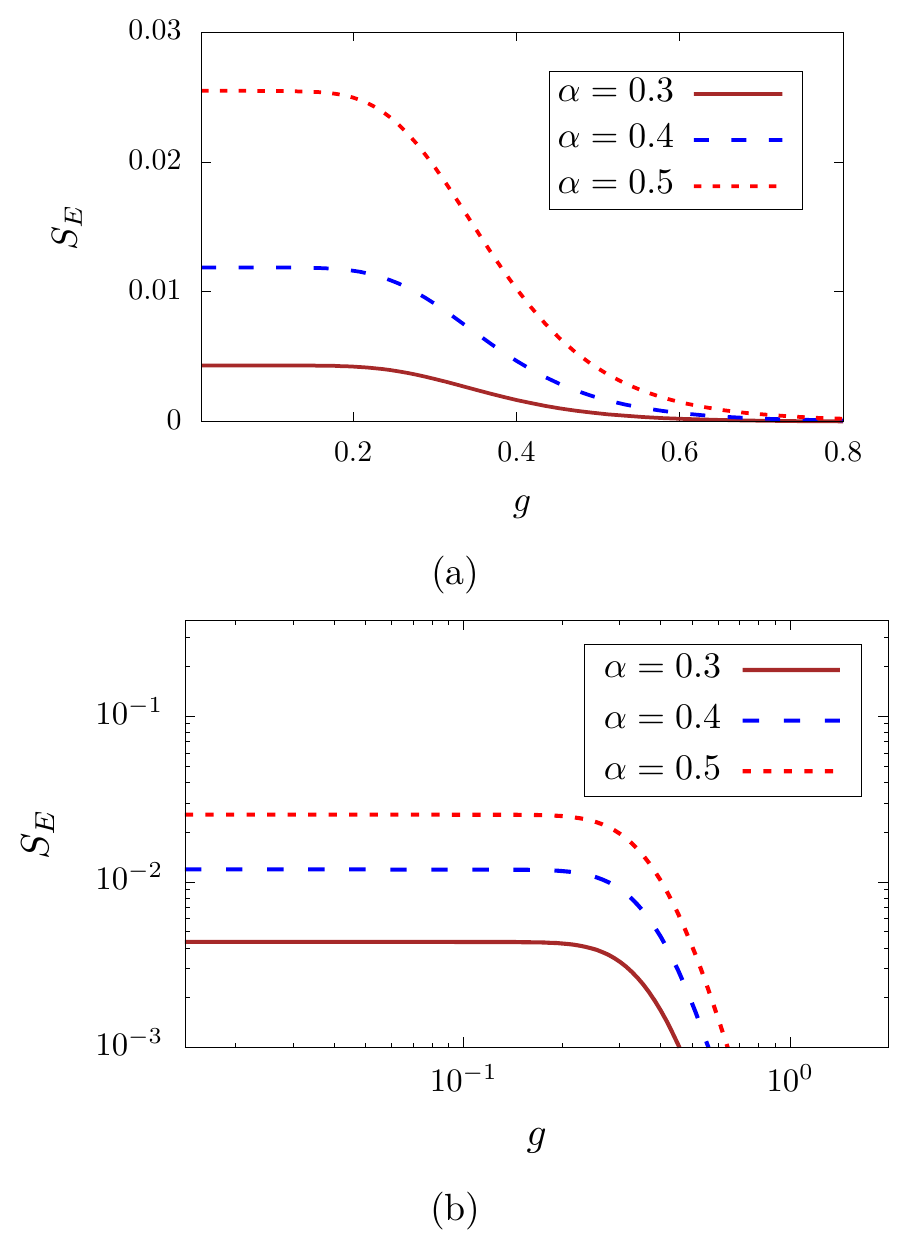}
  \caption{(a) The entanglement entropy of two coupled harmonic oscillators 
using perturbative formulation in polymer quantization, with varying $g$ keeping 
$\alpha=k_1/\omega$ fixed. (b) The $Log-Log$ plot of the entanglement entropy in 
polymer quantization, with respect to varying $g$ keeping $\alpha=k_1/\omega$ 
fixed.}
\label{fig:polymer-EE}
\end{figure}

We now intent to express this result in a more general fashion without using 
asymptotic forms of the Mathieu functions. For this we shall need some 
numerical help. First the expression of general energy difference of our 
concern in polymer quantization is
\begin{equation}
 E^{0}_{33}-E^{0}_{00} = \omega g[\mathcal{B}_{2}(g)-\mathcal{A}_{0}(g)]~, 
\label{eq:dEn_general_poly}
\end{equation}
where $\mathcal{A}_{n}(g)$ and $\mathcal{B}_{n}(g)$ are the Mathieu 
characteristic value functions corresponding to even and odd Mathieu functions. 
The expressions of $C_{03}$ are obtained from Eqn. (\ref{eq:Cn_general_poly}) 
with wave-functions represented in terms of Mathieu functions. Then we have
\begin{equation}\label{eq:perturbative_coefficient_poly}
 A=A_{33} = \frac{k_1^2}{\omega 
g[\mathcal{B}_{2}(g)-\mathcal{A}_{0}(g)]}C^1_{03}C^2_{03}~.
\end{equation}
and the corresponding eigen-values are
\begin{equation}
 \lambda_{1}=N_{1}^2=\left[1 + \left\{\frac{(\alpha 
~C_{03}/\lambda_{\star})^2}{\mathcal{B}_{2}(g)-\mathcal{A}_{0}(g)}\right\}^2 
\right]^{-1}~,
\end{equation}
and
\begin{equation}
 \lambda_{2} = \left\{N_{1}~\frac{(\alpha 
~C_{03}/\lambda_{\star})^2}{\mathcal{B}_{2}(g)-\mathcal{A}_{0}(g)}\right\}^2 ~,
\end{equation}
where $\alpha =k_1/\omega$. We have plotted this entanglement entropy coming 
from first order perturbation for different fixed values of $\alpha$ with 
varying polymer parameter $g$, in FIG. \ref{fig:polymer-EE}. Then the change of 
$g$ signifies the change in harmonic oscillator frequency $\omega$ for the fixed 
ratio $k_1/\omega$ and fixed polymer length scale $\lambda_{\star}$. In these 
plots the fixed ratio $(k_1/\omega)$ is considered to be less than one. It 
implies that the interaction strength is less than one and permits the 
application of perturbation theory even for high $g$ regimes. We observe that at 
high frequency regime as $g$ increases the entanglement entropy decreases and 
becomes very low at large $g$, see FIG. \ref{fig:polymer-EE}. This situation was 
not present in Schrodinger quantization as there the entanglement entropy is a 
function of $A=k_1^2/4\omega^2$ and its value is fixed for fixed $\alpha 
=k_1/\omega$. We want to note that same phenomena can also be observed in 
polymer quantization using second order perturbation with very little 
quantitative difference. For a unit mass harmonic oscillator one can interpret 
the inverse square-root of it's frequency to be a length scale characteristic of 
the harmonic oscillator. Now as the frequency increases this length decreases 
and even reaches polymer length scale $\lambda_{\star}$ when $g$ becomes very 
high. One then interprets the above deviation of entanglement entropy in polymer 
quantization from usual quantization, as a result of the physics in very high 
energy or in a very small length scale addressed by polymer quantization.

\subsection{Area law for free scalar field}

In this part we are going to use perturbation technique to evaluate the 
entanglement entropy corresponding to a massive free scalar field described by 
the Hamiltonian (\ref{eq:Field-hamiltonian-massive}). We first provide a 
prescription for the eigen-values of the reduced density matrix 
corresponding to $N-$coupled oscillators. By considering only first order 
perturbation, one can express the ground state wave-function of $N$ weakly 
coupled harmonic oscillators as
\begin{eqnarray}
|\Omega\rangle_N &=& |000...\rangle+A_1|nn0...\rangle+ 
...+A_{N-1}|...0nn\rangle,\nonumber\\\label{eq:Gen_perturb_waveFn}
\end{eqnarray}
where $A_1,...,A_{N-1}$ are coefficients of the first order perturbative 
correction to wave-function and they are functions of the individual frequency 
and interaction between different oscillators. In both Fock and polymer 
quantization the state $|00...n_{j}n_{j+1}...\rangle$ is obtained from 
$\hat{x}_{j}\hat{x}_{j+1}|00...\rangle=C^j_{0n} C^{j+1}_{0n} |00...n_{j}n_{j+1} 
...\rangle$, where we have omitted the sum on $n$ as the most dominating 
contribution comes from a single term. We first include appropriate 
normalization factor to the wave function from Eqn. 
(\ref{eq:Gen_perturb_waveFn}) and calculate the corresponding density matrix and 
its eigen-values for successively increasing number of coupled harmonic 
oscillators. Then the eigen-values, corresponding to the reduced density matrix 
after tracing over $\mathbbm{n}-$degrees of freedom out of total $N-$coupled 
harmonic oscillators, can be found by guessing from these consecutive 
eigen-value evaluation, as
\begin{eqnarray}
 \lambda^{\mathbbm{n}}_{s} &=& N^{2}_{c}A^2_{\mathbbm{n}}, 
~\frac{N^{2}_{c}}{2}\left[\sum_{\substack{j=0 \\ 
j\ne \mathbbm{n}}}^{N-1}A^2_{j}\pm \left\{\left(\sum_{\substack{j=0 \\ j\ne 
\mathbbm{n}}}^{N-1}A^2_{j}\right)^2 
\right.\right.\nonumber\\
~&& ~~~~~~~\left.\left. -~~4\left(\sum_{\substack{j=1 \\ j\ne 
\mathbbm{n}}}^{\mathbbm{n}-1}A^2_{j} 
\right)\left(\sum_{\substack{j=\mathbbm{n}+1 \\ j\ne \mathbbm{n}}}^{N-1}A^2_{j} 
\right)\right\}^{\frac{1}{2}} 
\right],\label{eq:eigenvalues-Ncoupled-perturbed-gen}
\end{eqnarray}
where $A_{0}=1$ and subscript $s$ denotes different eigenvalues which are three 
in number for any particular reduction $\mathbbm{n}$. $N_{c} = (\sum_{j=0}^{N-1} 
A^{2}_{j})^{-1/2}$ is the normalization factor corresponding to the perturbed 
ground state. As discussed earlier the entanglement entropy corresponding to 
these eigenvalues would be $S_{E}^{\mathbbm{n}} = -\sum_{s=1}^{3} 
\lambda^{\mathbbm{n}}_s \ln{\lambda^{\mathbbm{n}}_s}$.

\subsubsection{Area law in Fock quantization}


\begin{figure}
  \includegraphics[width=0.8\linewidth]{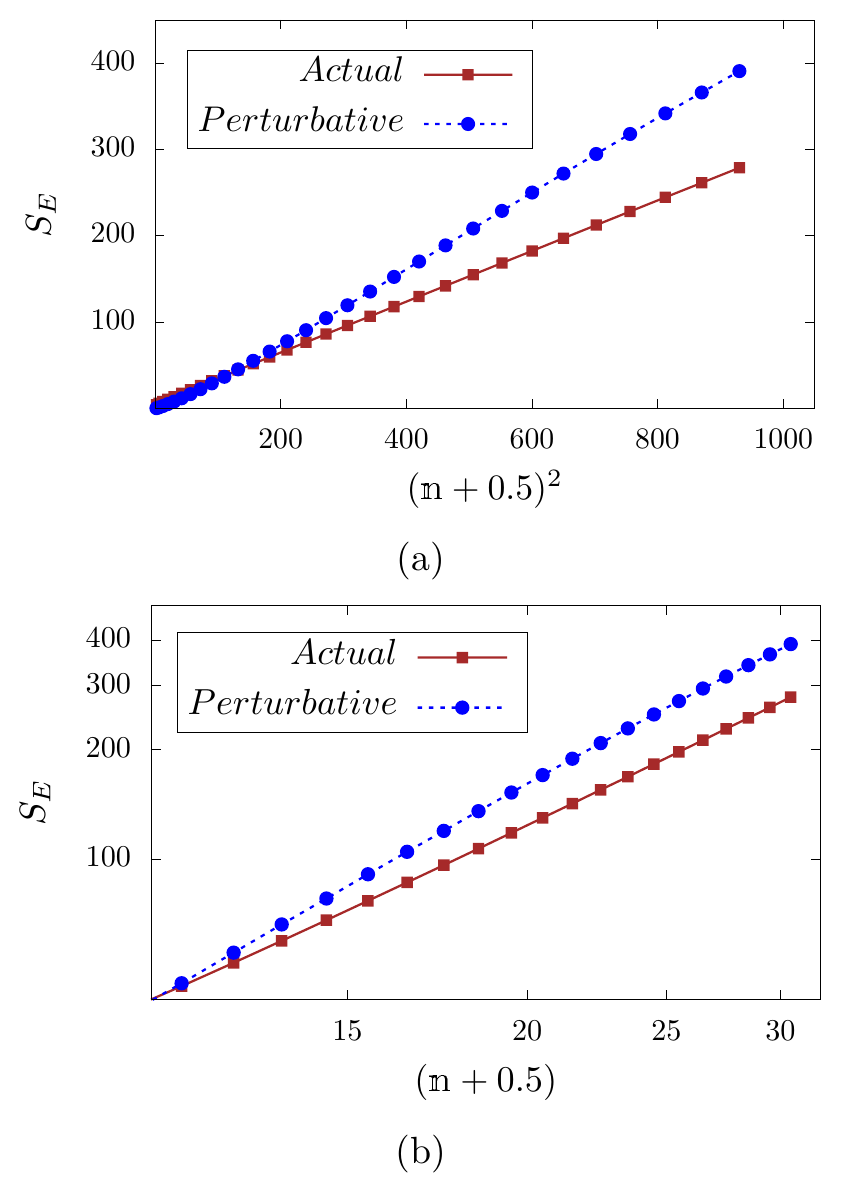} 
  \caption{(a) The plot of entanglement entropy with respect to 
$(\mathbbm{n}+0.5)^2$ for a free massless scalar field in a lattice of total 
size $N=60$. Here $\mathbbm{n}$ is the number of position degrees of freedom 
traced out of the ground state of the scalar field. This plot shows the area law 
for entanglement entropy in both actual and perturbative formulation. Here the 
ratio of the slopes from perturbative entropy and actual entropy is $\sim 1.85$. 
(b) This figure provides the $Log-Log$ plot of entanglement entropy vs 
$(\mathbbm{n}+0.5)$ in both actual and perturbative formulation. The slope of 
the curve from actual formulation is about $\sim1.9$ where the slope from 
perturbative formulation is $\sim2.1$. We have used first order perturbation to 
evaluate the entanglement entropy.}
  \label{fig:area-law-comparison}
\end{figure}

In order to obtain the area law of entanglement entropy in Fock quantization we 
first consider the discrete Hamiltonian, formed out of partially Fourier 
transformed field Hamiltonian in a lattice of finite size, from Eqn. 
(\ref{eq:Hamiltonian-massive-FT-discrete1}). With the help of Eqn. 
(\ref{eq:Interaction_matrix_discrete_hamiltonian}) one can get frequency 
$\omega_j$ of the $j^{th}$ oscillator and coupling $k^2_j$ between $j^{th}$ and 
$(j+1)^{th}$ oscillator. Now according to Eqn. 
(\ref{eq:perturbative_coefficient}) we want to find the expressions of the 
coefficients of first order perturbation $A_{j} = k^2_{j}/ 
\left\{2\sqrt{\omega_{j} \omega_{j+1}} (\omega_{j} + \omega_{j+1}) \right\}$, 
which are used in Eqn. (\ref{eq:eigenvalues-Ncoupled-perturbed-gen}) to get the 
eigen-values. We note that the perturbative coefficients $A_{j}$ are in 
principle functions of $l$ and $m$ and sums over these quantities are taken to 
evaluate the entropy, but for brevity we omitted their index from the notation. 
In Fock quantization they are evaluated using $C^j_{0n} = 
\delta_{1,n}/\sqrt{2\omega_j}$ and $E^j_n=(n+1/2)\omega_j$. We have numerically 
computed the entanglement entropy using the obtained eigen-values. In FIG. 
\ref{fig:area-law-comparison} the entanglement entropy from perturbative and 
actual formulation are presented for a massless free scalar field.
The entanglement entropy in actual formulation is obtained non perturbatively, 
utilizing the Gaussian nature of the ground state wave-functions. We have used 
the results from Eqn. (\ref{eq:rho_nreduced}) and the potential from Eqn. 
(\ref{eq:Interaction_matrix_discrete_hamiltonian}) to evaluate the entanglement 
entropy in actual formulation \cite{Srednicki:1993im}.
FIG. \ref{fig:area-law-comparison} shows that first order perturbation is 
sufficient enough to provide an area law for entanglement entropy. We want to 
note that the slope from this area curve($\sim 0.42$) is different than the one 
obtained from actual formulation($\sim 0.29$). 
One can notice from Eqn. (\ref{eq:Hamiltonian-massive-FT-discrete1}, 
\ref{eq:Interaction_matrix_discrete_hamiltonian}) that the ratio of the 
frequency square to interaction strength decreases when $j$ increases as we 
consider a larger system size. Thus perturbation theory becomes less effective 
and the results obtained from first order perturbation deviates more from the 
actual non perturbative result for large $j$.
We also want to note that as second order perturbation is employed in this 
formulation the area curve is quantitatively very little improved and gets 
closer to the area curve from actual formulation. Now as we consider a field 
with increasing mass, the ratio of the frequency square and coupling strength 
from Eqn. (\ref{eq:Interaction_matrix_discrete_hamiltonian}) increases and the 
perturbative formulation becomes more effective in describing the original 
system.


\begin{figure}
    \includegraphics[width=0.8\linewidth]{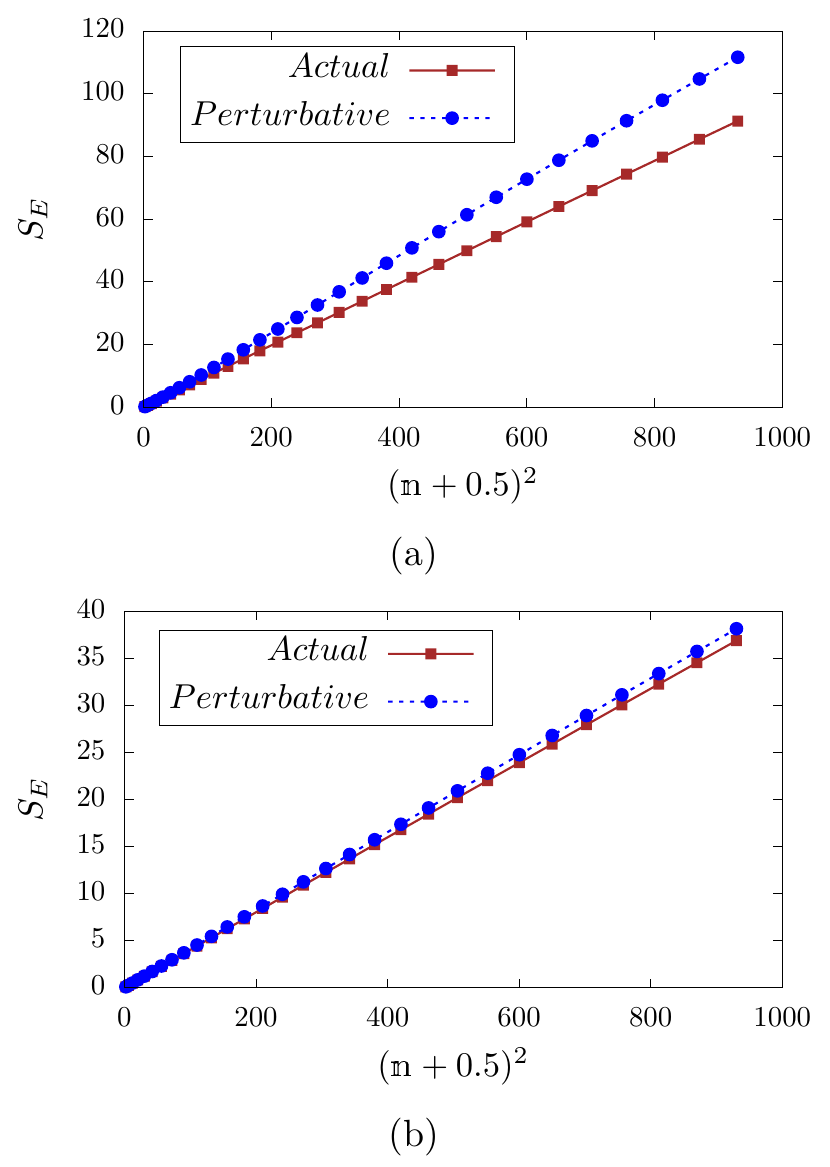}
    \caption{(a) The plot of entanglement entropy of a massive free scalar field 
with respect to $(\mathbbm{n}+0.5)^2$ for $N=60$. Here $\mu a=2$ and the ratio 
of the slopes from perturbative entropy and actual entropy is $\sim 1.22$. (b) 
The entanglement entropy vs $(\mathbbm{n}+0.5)^2$ plot for massive free scalar 
field with $N=60$. Here $\mu a=4$ and the ratio of the slopes from perturbative 
entropy and actual entropy is $\sim 1.04$. In both cases only first order 
perturbation is used.}
\label{fig:area-law-comparison-mu24}
\end{figure}

In addition to the massless case we have also considered massive scalar field 
with discretized Hamiltonian from Eqn. 
(\ref{eq:Hamiltonian-massive-FT-discrete1}). We have taken different values for 
the parameter $\mu a$ and for each value obtained the area law in both the 
actual and first order perturbation. Then the massless case becomes a special 
case when this parameter $\mu a$ is taken to be zero. In particular we have 
observed that as we increase the value of this parameter the perturbation 
becomes stronger and the slopes of the two area curves corresponding to actual 
and perturbative formulation get closer. The plots corresponding to this 
feature are shown in FIG. \ref{fig:area-law-comparison-mu24}.

\subsubsection{Area law in polymer quantization}

We take the Fourier Hamiltonian density for a massive free scalar field as 
according to equation (\ref{eq:Hamiltonian-massive-FT-discrete1}). We note that 
the field modes $\varphi_{lm,j}$ and their conjugate momentums $\pi_{lm,j}$ are 
dimensionless here. To consistently introduce polymer quantization in this 
formulation we make the transformations $\pit_{lm,j} \equiv \pi_{lm,j}/\sqrt{a}$ 
and $\varphit_{lm,j} \equiv \sqrt{a}~\varphi_{lm,j}$ such that the new field 
mode and the conjugate momentum become dimension full. Then the Fourier 
Hamiltonian density from Eqn. (\ref{eq:Hamiltonian-massive-FT-discrete1}) will 
take the form
 \begin{eqnarray}
H_{lm} &=&  \tfrac{1}{2} \sum_{j=1}^{N} \left[ 
\pit^2_{lm,j}+\left(\frac{j+\frac{1}{2}}{a}\right)^2\left(\frac{\varphit_{lm,j
} }
{j}
-\frac{\varphit_ {lm,j+1}}{j+1} 
\right)^2 \right.\nonumber\\
~&& ~~~~~~~~~~~~~\left. +\left\{\frac{l(l+1)}{a^2j^2}+\mu^2\right\}\varphit^2_{
lm,j} \right]~.\label{eq:fourier-hamiltonian-field-discrete3}
\end{eqnarray}
This Hamiltonian also describes a system of $N$ coupled harmonic oscillators 
given by Eqn. (\ref{eq:general-Ncoupled-HO-hamiltonian}). In polymer 
quantization a new dimension full parameter $\lambda_{\star}$ is introduced with 
dimension $(length)^{1/2}$, inverse of the dimension of momentum. Here the basic 
variables are taken to be $\varphit_{lm,j}$ and $U^{\lambda_{\star}}_{lm,j} = 
\exp{\left\{i\lambda_{\star} \pit_{lm,j}\right\}}$ with Poisson bracket 
$\{\varphit_{lm,j},U^{\lambda_{\star}}_{lm,j}\} = i\lambda_{\star} 
U^{\lambda_{\star}}_{lm,j}$. From the above system of coupled harmonic 
oscillators we observe that for a general $j^{th}$ oscillator the frequency is
\begin{eqnarray}
 \omega_{lm,j} &=& \frac{1}{a~j}\left[l(l+1)+\left\{\left(j+\tfrac{1}{2} 
\right)^2+\left(j-\tfrac{1}{2}\right)^2\right\}\right|_{j\neq1,N} 
\nonumber\\
~&& \left.+\mu^2 a^2 j^2+\tfrac{9}{4}\delta_{j1} + 
\left(N-\tfrac{1}{2}\right)^2\delta_{jN}\right]^{\frac{1}{ 2 } }\nonumber\\ 
~&=& \frac{\Omega_{lm,j}}{a}~.
\end{eqnarray}
%

\begin{figure}
  \includegraphics[width=0.8\linewidth]{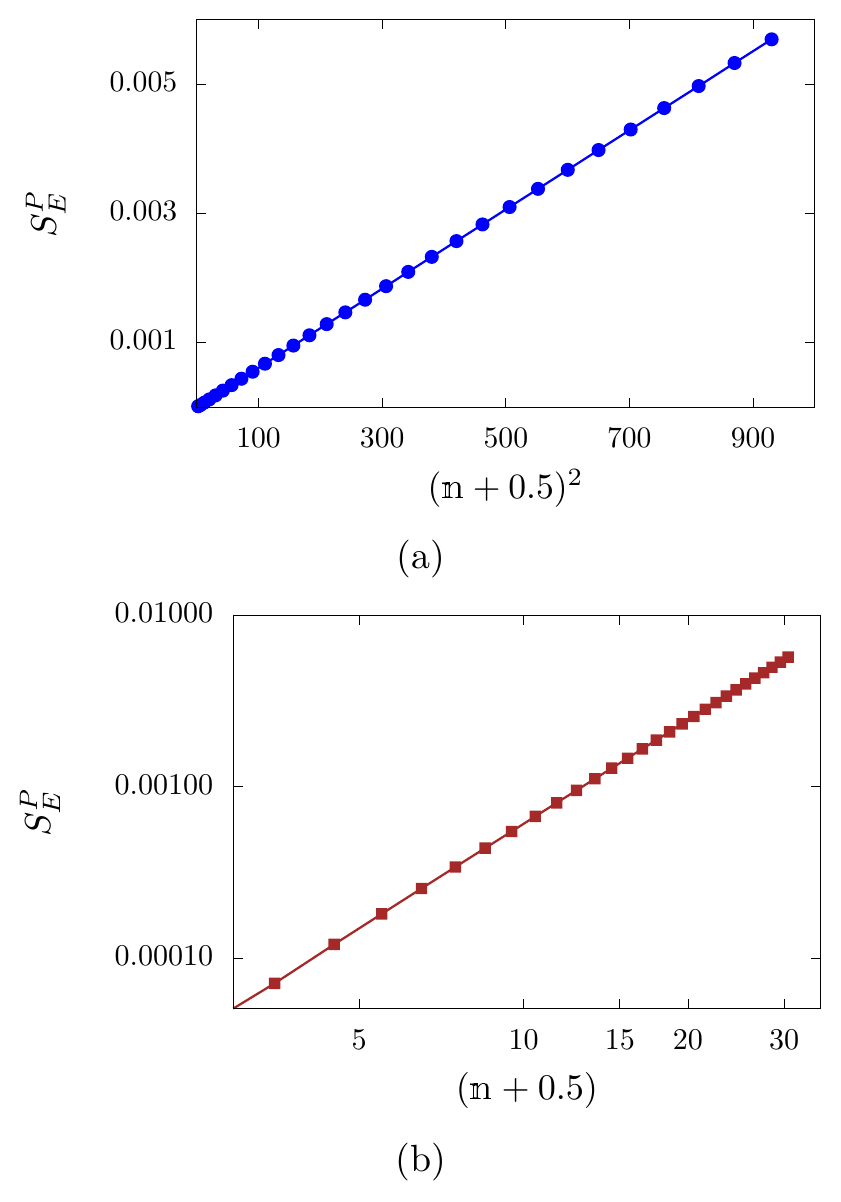} 
  \caption{(a) The entanglement entropy of a massless free scalar field with 
respect to $(\mathbbm{n}+0.5)^2$ using first order perturbation in polymer 
quantization. This plot shows the area law in polymer quantization with slope 
$6.12\times10^{-6}$. (b) The $Log-Log$ plot of the entanglement entropy vs 
$(\mathbbm{n}+0.5)$ in perturbative formulation using polymer quantization. The 
total lattice size is taken to be $N=60$ and the obtained slope is $2.002$.}
  \label{fig:Poly-area-law}
\end{figure}

Now we want to get the expressions of perturbative coefficients $A_{j}$  used in 
Eqn. (\ref{eq:eigenvalues-Ncoupled-perturbed-gen}). They are constructed using 
expressions of $C^j_{0n}$ and $E^j_n$ in polymer quantization from Eqn. 
(\ref{eq:Cn_general_poly}) and (\ref{eq:dEn_general_poly}), which are further 
given by the dimension less polymer parameter $g_{lm,j} = \omega_{lm,j} 
\lambda^2_{\star} = \Omega_{lm,j}l_{\star}/a$. We also want to note that the 
inter atomic distance $a$ and the polymer length scale 
$l_{\star}=\lambda^2_{\star}$ both have same nature and should have same order 
as they signify the ultraviolet cutoff. Then we take their ratio 
$\gamma=l_{\star}/a$ to be unity, which adds further simplification to the 
evaluation of entanglement entropy. We have plotted the entanglement entropy 
from first order perturbative formulation in FIG. \ref{fig:Poly-area-law} 
considering massless free scalar fields in polymer quantum field theory. In 
these figures we observe that the area law is valid in polymer quantization too. 
However the corresponding slope is now very low compared to results from Fock 
quantization. One can also get the area law in polymer quantization for massive 
free scalar field with a further decreased slope.

\begin{figure}
  \includegraphics[width=0.8\linewidth]{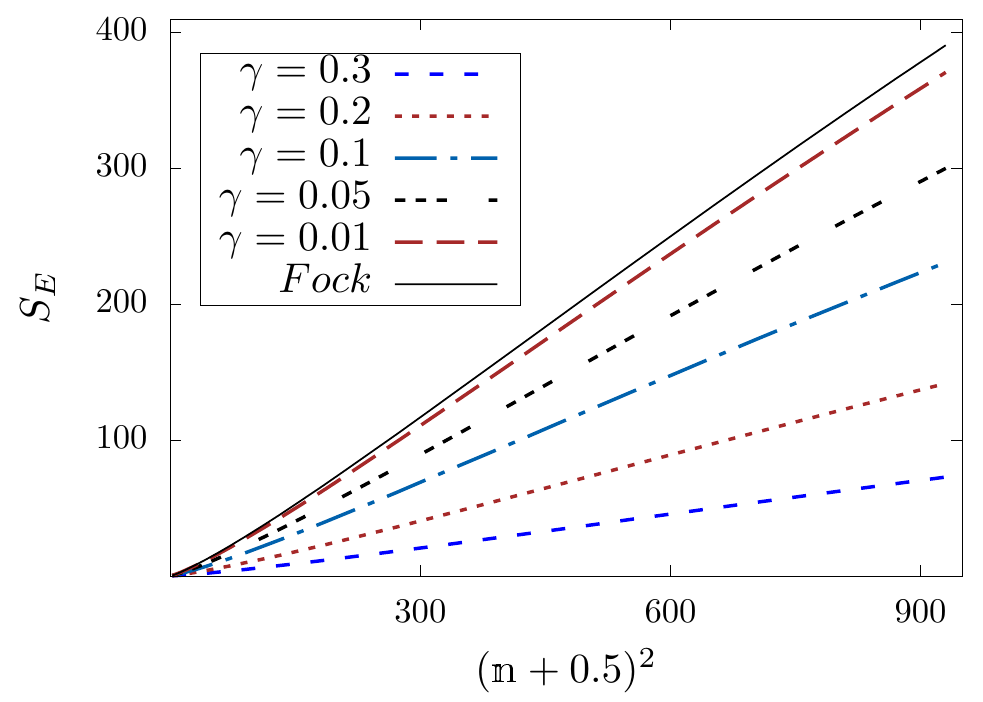}
  \caption{The plot of entanglement entropy in polymer quantization for 
different values of $\gamma=l_{\star}/a$.}
  \label{fig:Poly_Slope_Change}
\end{figure}

\begin{figure}
  \includegraphics[width=0.8\linewidth]{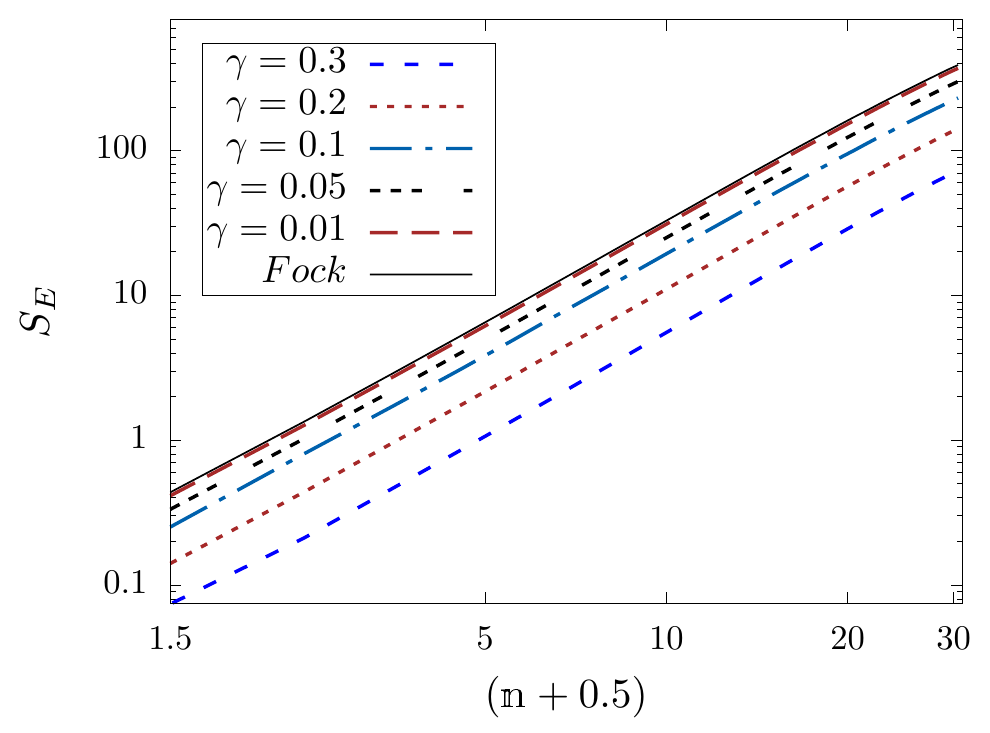}
  \caption{The $Log-Log$ plot of entanglement entropy in polymer quantization 
for different values of $\gamma=l_{\star}/a$.}
  \label{fig:log_Poly_Slope_Change}
\end{figure}

\emph{Implication of the result:}  
From \cite{Das:2008sy} we get to understand that the slope of the area curve for 
entanglement entropy can be different due to many reasons, such as due to 
different discretization procedures, inclusion of mass or taking excited states 
instead  of the ground state. We want to mention here a consistency check to 
understand whether this result from polymer quantization is a plausible one or 
not. It is noted in \cite{Hossain:2010eb} that in low energy regimes polymer 
quantization reproduces the results from usual Fock quantization. Now in this 
formulation of entanglement entropy evaluation we observe that one direct 
influence of polymer quantization over Fock is dictated by the factor 
$(l_{\star}/a)$. When this factor is unity the system is completely interpreted 
in terms of polymer quantization. On the other hand when the value of this 
quantity decreases the value of the dimensionless polymer frequency $g_{lm,j}$ 
decreases and the system becomes more and more Fock like as the lower energy 
regimes of polymer quantization tends to contribute to the description of the 
system. We have plotted the entanglement entropy for different values of this 
factor and we observed that as the value decreases the area curve of 
entanglement entropy from polymer quantization approaches the one from Fock 
quantization, see FIG. \ref{fig:Poly_Slope_Change} and FIG. 
\ref{fig:log_Poly_Slope_Change}. Thus the very low slope of the entanglement 
entropy can be described as a feature coming from the  disentangling nature of 
polymer quantization at high energy regimes. We want to note that massive scalar 
fields also show disentangling nature and lowers the slope of the area curve 
\cite{Katsinis:2017qzh, Riera:2006vj, Balasubramanian:2011wt}.

The entanglement entropy of free scalar field in polymer quantization gives rise 
to another question, which relates to the corrections to the area law as 
predicted by quantum gravity \cite{Pasqua:2014hea, Sen:2012cj, 
El-Menoufi:2015cqw, Pathak:2016vfc}. In this manner we want to note that the 
slope of the entanglement entropy vs $(\mathbbm{n}+0.5)$ curve in \emph{Log-Log} 
plot is $2.002$, which automatically discards any possible departure from the 
area law. This area dependence of entanglement entropy in polymer quantization 
is enthralling in its own right since it validates the generality of the area 
law in quantizations other than Fock.

\section{Discussion}

In usual formulation, procurement of the area curve for entanglement entropy 
\cite{Bekenstein:1974ax, Kallosh:1992wa, hawking1975, Bekenstein:1975tw, 
Hawking:1976de, Gibbons:1976pt, Davies:1978mf} is simplified using the 
mathematical structure of Gaussian ground state wave-function from Schrodinger 
quantization. However not all quantization procedures provide this Gaussian 
nature of ground state and polymer quantization is one of them.
We note that though entanglement entropy for two coupled harmonic oscillators 
are specifically evaluated for polymer quantization in \cite{Demarie:2012tz}, 
the framework to obtain entanglement entropy for large number of coupled 
oscillators is not provided thus one can not obtain the area law. In this work 
we have treated the interaction between coupled harmonic oscillators in 
perturbative manner. Our procedure is different than the ones discussed in 
\cite{Katsinis:2017qzh, Riera:2006vj, Balasubramanian:2011wt, Kumar:2017ctm}, 
where the eigen-values of the reduced density matrix and momentum space 
entanglement are estimated using perturbation.
For two coupled harmonic oscillators we noticed disentangling nature from 
polymer quantization at high frequency regime. We observed that in Schrodinger 
quantization the entanglement entropy is unchanged while in polymer quantization 
it decreases at high oscillator frequencies, keeping the ratio of interaction 
strength to frequency square fixed. We showed that in our formulation, by 
considering free scalar field, one obtains the area law of entanglement entropy 
for Fock quantization. As the mass of the scalar field increases the individual 
oscillator frequency increases, thus perturbation strength increases and the 
obtained area curve approaches the area curve from usual formulation. 
Furthermore we showed that in polymer quantization also this formulation 
provides a similar area law, but with a very decreased slope. We inferred that 
this decrease of slope is due to the disentangling nature of polymer 
quantization at higher energies. We further noticed that as the effect of 
polymer quantization becomes smaller, by lowering the value of the ratio of 
polymer length scale $l_{\star}$ to inter-atomic distance $a$, the area curve 
from polymer quantization using first order perturbation tends to approach the 
area curve from Fock quantization. This phenomena is not quite surprising as in 
the limit $l_{\star}/a \rightarrow 0$, the physical result from polymer 
quantization would converge to the result obtained from the standard Fock 
quantization.
The disentangling nature of polymer quantization is very intriguing in its own 
right as it is known that usual quantization looses its predictability in 
trans-Planckian energy regimes \cite{Cohen:1998zx, Carmona:2000gd}.
We note that this disentangling phenomena in polymer quantization is analogous 
to the suppression of propagation at large energies.
We mention that there are other derivations to obtain the area law and harvest 
entanglement entropy for scalar field \cite{Calabrese:2004eu, Callan:1994py, 
Calabrese:2005zw, Solodukhin:2011gn, VanRaamsdonk:2009ar, Casini:2009sr, 
Cramer:2005mx, Page:1983ug, Kabat:1994vj, Plenio:2004he, Solodukhin:1994yz, 
Holzhey:1994we, Das:2001ic, Allouche:2018err, Pagani:2018mke, Henderson:2017yuv, 
Reznik:2002fz} and it would be interesting to see whether an exact form of the 
entanglement entropy can be found using these derivations in polymer 
quantization. In conclusion we address that our formulation opens up an avenue 
to understand entanglement entropy in terms of perturbative corrections.\\

\begin{acknowledgments}
We would like to thank Golam Mortuza Hossain, Narayan Banerjee, Ritesh K. Singh 
and Ananda Dasgupta for discussions. We would also like to thank Sumanta 
Chakraborty, Abhishek Majhi and Sudipta Saha for useful suggestions. S.B. would 
like to thank IISER Kolkata for supporting this work through a doctoral 
fellowship.
\end{acknowledgments}



\end{document}